# Poole-Frenkel Effect and Phonon-Assisted Tunneling in GaAs Nanowires

*Aaron M. Katzenmeyer[1], François Léonard[1*], A. Alec Talin[1†], Ping-Show Wong[2], Diana L. Huffaker[2]*

[1]Sandia National Laboratories, Livermore, CA 94551

[2]Electrical Engineering Department, University of California at Los Angeles, Los Angeles, CA 90095

[*]fleonar@sandia.gov

[†]Currently at Center for Nanoscience and Technology, NIST, Gaithersburg, MD

We present electronic transport measurements of GaAs nanowires grown by catalyst-free metal-organic chemical vapor deposition. Despite the nanowires being doped with a relatively high concentration of substitutional impurities, we find them inordinately resistive. By measuring sufficiently high aspect-ratio nanowires individually *in situ*, we decouple the role of the contacts and show that this semi-insulating electrical behavior is the result of trap-mediated carrier transport. We observe Poole-Frenkel transport that crosses over to phonon-assisted tunneling at higher fields, with a tunneling time found to depend predominantly on fundamental physical constants as predicted by theory. By using *in situ* electron beam irradiation of individual nanowires we probe the nanowire electronic transport when free carriers are made available, thus revealing the nature of the contacts.

KEYWORDS EBIC, electroconductivity, GaAs, phonon-assisted tunneling, Poole-Frenkel effect, nanowire

Due to the variety of chemical composition and numerous growth techniques, a detailed understanding of the fundamental electrical transport properties of semiconducting nanowires (NWs) remains in its infancy. Such a knowledge base is necessary however for nanowires to serve in any substantive capacity in future technologies. A major task is identifying the particular charge transport phenomena that dominate in a given nanowire system. This is important not only to take advantage of the transport physics for device engineering, but also to correlate electronic transport with basic materials issues. For example, the nanowire growth conditions can significantly impact the material quality through the introduction of dopants, impurities and defects; and growth also determines the nanowire shape which directly influences transport through, e.g. surface states. Measurement and theory have recently revealed several different transport regimes and effects. Examples of this include Ohmic transport for Si NWs,[1-3] space-charge-limited current in GaN,[4] InAs,[5] CdS[6] and trap-containing GaAs NWs,[7] recombination current in Au-catalyst/Ge NW Schottky diodes,[8] ballistic and diffusive transport in InAs NWs[9] and size-dependent mobility[5, 10] and carrier concentration[5] in InAs NWs. Here we show that yet other transport regimes, Poole-Frenkel and phonon-assisted tunneling, arise in GaAs NWs by utilizing *in situ* nanoprobing of individual nanowires directly on the growth substrate. These regimes arise due to the presence of traps along the length of the NWs, as confirmed by a novel characterization technique.

The vertically aligned NWs used in this study were grown by catalyst-free metal-organic chemical vapor deposition (MOCVD) on a degenerately doped (*n*++) GaAs (111)B wafer coated with a $SiO_2$ template. E-beam lithography was used to create arrays of holes in the $SiO_2$, which determined NW diameter. The wires were grown at 700°C for 10 minutes with a V-III ratio of 10, using *tert*-butylarsine and trimethylgallium precursors in hydrogen carrier gas at a total reactor pressure of 60 Torr. The nanowires were doped n-type with Si; a bulk sample grown under these conditions contained a Si concentration of $7 \cdot 10^{18}/cm^3$, and exhibited linear current-voltage characteristics with a resistivity of $\sim 10^3$ A/V-cm.

Electrical measurements on individual NWs were performed *in situ* in a field-emission scanning electron microscope (FE-SEM) retrofitted with a nanoprobe system. We have reviewed this measurement technique in detail elsewhere.[11] Briefly, a high resolution electronic manipulator is fitted with a tungsten (W) probe tip which has been electrochemically etched to nanoscopic dimension. This probe contacts a single NW at its tip, serving as one electrode; the other electrode is provided by the degenerately doped growth substrate which is contacted by the substrate holder via a large area indium back-contact. In this way we are able to measure NWs which are free of perturbative influences present whenever NWs are measured in ambient or whilst lying on a substrate.

Figure 1a depicts a typical measurement and figure 1b shows a representative I-V characteristic for NWs in this study. Attention is called to the high turn-on voltage of several tens of volts and its symmetry between polarities. This behavior is indicative of electronic transport limited by processes along the length of the NW, as opposed to contact effects, since the latter are asymmetric. Indeed one expects an Ohmic contact at the substrate/NW interface given dopant type and concentration and a Schottky contact at the NW/W probe interface given that defects commonly pin metal Fermi levels near mid-bandgap in GaAs.[12] Thus, if the transport was injection-limited, a strong asymmetry in the I-V characteristic between polarities would be observed with a turn on voltage at forward bias given by $E_g/2e \approx 0.72V$, where $E_g$ is the bandgap of GaAs. The fact that the data does not follow this picture implies that the transport is dominated by the bulk of the nanowire. (We show below that the *injection-limited* nature of the system, once revealed, is consistent with this picture.)

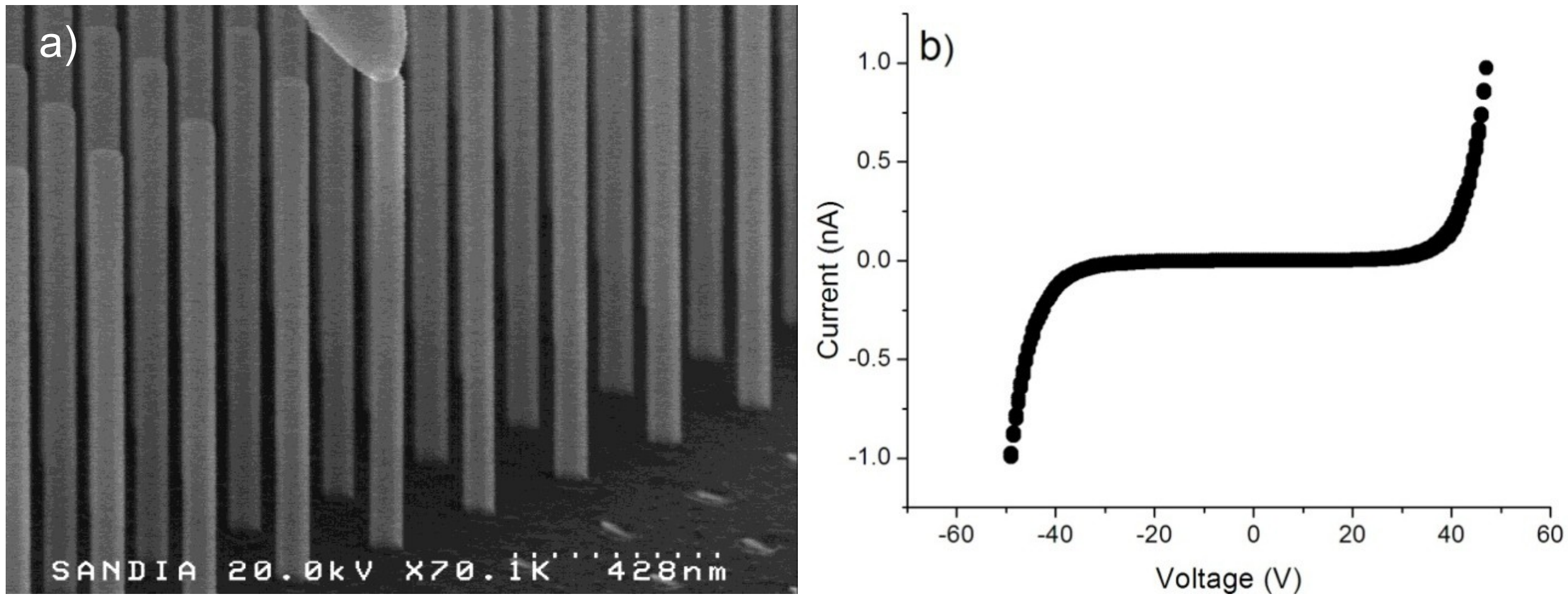

**Figure 1: a)** FE-SEM micrograph depicting a typical *in situ* electrical measurement. **b)** A representative I-V curve. The e-beam is directed away from the NW during measurement unless otherwise specified.

The large electric fields required to observe relatively small currents are quantitatively consistent with concepts of trap-dominated transport. Indeed Frenkel showed that in a semiconductor or insulator, carrier traps, which are charged when empty and neutral when filled, can mediate carrier transport and give rise to a pre-breakdown phenomenon in which the increase in conductivity is exponentially proportional to the square root of the electric field.[13] This relationship arises as the applied field lowers the barrier of the trap Coulombic potential, thus enhancing thermionic emission over the trap barrier. This model has been reworked in order to account for experimental results;[14-15] however, we find that our GaAs NWs are well described by the classical, one-dimensional model in which:[14]

$$\sigma \propto \exp\left(\frac{e^3 E}{\pi\varepsilon (kT)^2}\right)^{1/2},$$

where $\sigma$ is the conductivity, $E$ is the electric field, $\varepsilon$ is the dielectric constant, $k$ is the Boltzmann constant, and $T$ is the temperature. Figure 2a shows data for the NWs plotted at forward and reverse bias under this formalism along with linear fits. (The coefficient of determination in each instance was ≥ 99.3%. Fits to other power laws gave nonphysical values of ε.)

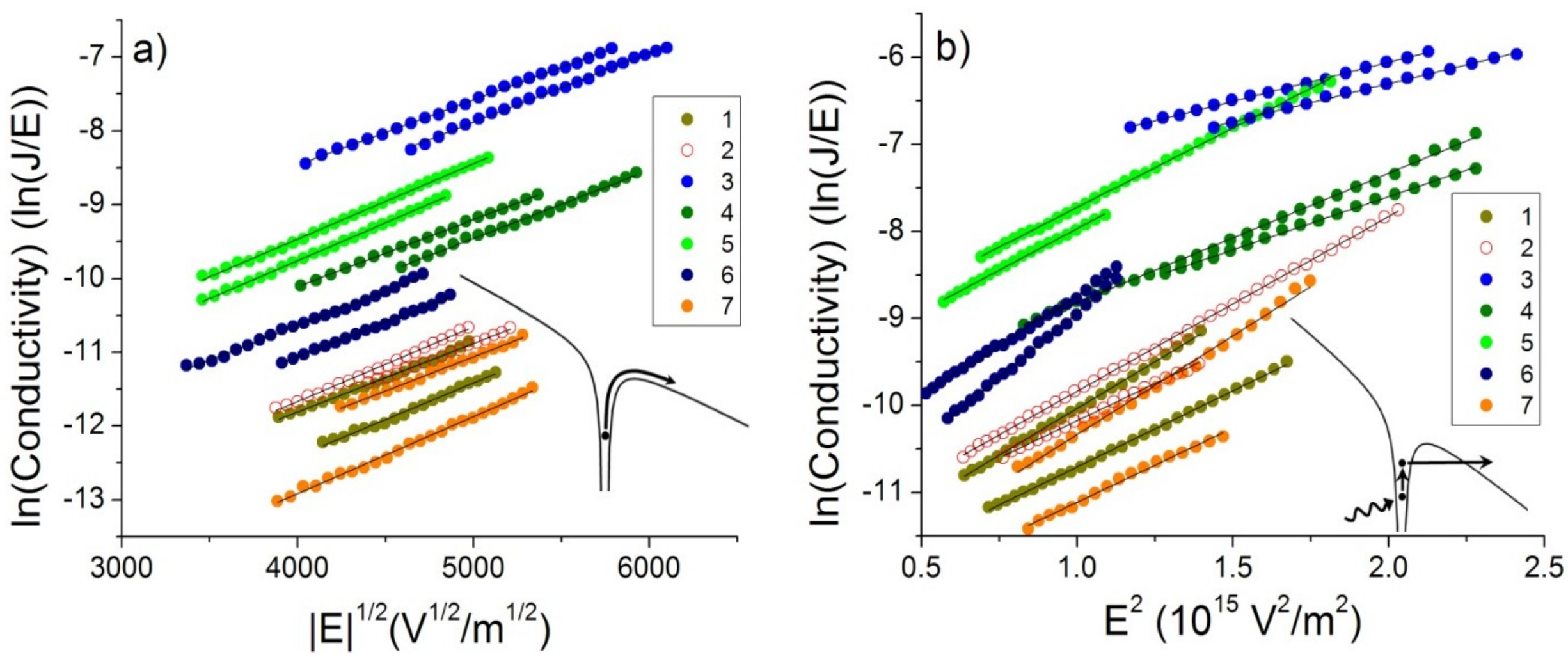

**Figure 2: a)** Intermediate-field and **b)** high-field data plotted respectively in the formalisms of Poole-Frenkel effect and phonon-assisted tunneling. The insets schematically depict carrier escape from the trap in each regime. The voltage range over which a transport mechanism was observed in a given NW can be obtained using the NW lengths in Table 1.

The linearization of the data indicates that the conductivity is indeed exponentially dependent on $E^{1/2}$ in this intermediate field regime. Schottky emission would also be linearized in such a plot, so care must be taken in ascribing the appropriate mechanism in the absence of other indicators such as that discussed above. Discrimination between the two effects is provided by analyzing the slope of the lines (lower by a factor of 2 for Schottky emission) which typically enables one to extract a sensible value for the dielectric constant under the correct formulation.[15] Using the Poole-Frenkel formulation we find the nanowires to have dielectric constants near that of bulk GaAs which further affirms bulk-limited transport and the linear fits in figure 2. The dielectric constant values extracted for all wires in this study at forward and reverse bias are given in Table 1 as are the NW dimensions.

Table 1: NW Characterization Values

| Wire | R (nm) | L (μm) | $\varepsilon_{r+}$ | $\varepsilon_{r-}$ | $\tau_{+}$ (fsec) | $\tau_{-}$ (fsec) |
|---|---|---|---|---|---|---|
| **1** | 33 | 0.99 | 10.7 | 9.08 | 11.7 | 11.1 |
| **2** | 40 | 1.032 | 8.83 | 10.6 | 11.6 | 10.9 |
| **3** | 42 | 0.672 | 11.6 | 10.2 | 8.77 | 8.68 |
| **4** | 38 | 0.712 | 10.5 | 10.5 | 10.4 | 9.71 |
| **5** | 41 | 1.05 | 8.65 | 8.48 | 11.1 | 11.3 |
| **6** | 42 | 1.013 | 10.1 | 10.3 | 13 | 12 |
| **7** | 43 | 0.865 | 10.5 | 8.32 | 12 | 10.9 |

With application of increasingly higher electric field, a competitive transport mechanism emerges as a result of field-induced thinning of the trap energy barrier. Rather than being fully emitted over the barrier, it becomes more probable for a carrier to absorb a lesser amount of energy from phonons and subsequently tunnel through the trap barrier. This mechanism, aptly named phonon-assisted tunneling results in a conductivity increase described by:[16]

$$\sigma \propto \exp\left(\frac{(eE)^2 \tau^3}{3m^* \hbar}\right),$$

where $\tau$ is the tunneling time and $m^*$ the effective carrier mass. The hallmark of this transport mechanism is that conductivity increases exponentially with the square of the applied electric field. Increasing the applied field beyond values where Poole-Frenkel transport is applicable, in all cases resulted in plots that become superlinear in the Poole-Frenkel formulation. We plot this high field region of the datasets in the formalism of phonon-assisted tunneling as represented in figure 2b. Again, data from forward and reverse bias regions are shown. The data are well linearized, indicating that at relatively high fields carriers traverse the NW via phonon-assisted tunneling events. The slope of the line enables determination of the tunneling time and these values are listed for each wire in forward and reverse bias in table 1. (We used the GaAs bulk effective carrier mass of $0.067m_0$.) The tunneling time predicted from theory,

$$\tau = \frac{\hbar}{2kT} \pm \tau',$$

is increased or decreased from a value determined entirely by fundamental constants and reciprocal temperature by $\tau'$ depending on the nature of the traps in the material (i.e. whether the traps facilitate weak or strong electron-phonon coupling, respectively).[17] In all NWs we observe similar tunneling times which are near and less than $\hbar\ 2kT = 13$ femtoseconds, consistent with strong electron-phonon coupling and indicating the traps are autolocalized centers.[18] An affirmation and consequence of this are discussed below. (Determining the exact nature of the traps is beyond the scope of this paper, requiring detailed analysis using, e.g. Deep-Level Transient Spectroscopy. Two possible candidates are the EL2 level and DX centers, both of which have been determined to be self-trapping in GaAs.[19-22])

Finally to confirm that traps dominate the transport and to reveal the nature of the contacts we employ a novel characterization technique – electroconductivity measurements - in conjunction with the well known electron beam induced current (EBIC)[23] technique. In EBIC a focused electron beam is rastered over the semiconductor and the created electron-hole pairs which diffuse to a built-in potential (e.g. p-n junction or Schottky contact) are separated and used to generate a spatial current profile or an EBIC image in which the contrast directly corresponds to the collected current. Figure 3a shows a representative EBIC profile collected as the beam is rastered along the length of a NW. Net current is only collected in the vicinity of the built-in potential formed by the W contact. The trough (as opposed to a peak) in the profile indicates that a hole current was collected at the W tip (and an electron current at the substrate). The lack of net current collection when the beam is near the substrate/NW interface indicates Ohmic contact. Therefore the EBIC profile affirms the asymmetry of the contacts and the injection-limited nature of the system when traps are activated.

Complementarily, rather than rastering the beam with the device at fixed (zero or reverse) bias we focus it to a point on the NW and perform a I-V sweep. This technique allows us to probe the response of the system when free carriers are made available and quantitatively determine the Schottky barrier height (i.e. the applied voltage for which the net current is nullified). Figure 3b shows a I-V curve recorded during such an electroconductivity experiment. The I-V curve is strongly asymmetric, with a turn-on voltage at forward bias less than 1V, consistent with the contact being pinned at mid-bandgap. The measured beam current at the sample (10 pA at +70 V bias of the collector) cannot account for this substantial change in behavior during e-beam exposure; the increased current is the result of the creation of multiple carriers per injected e-beam electron.[23]

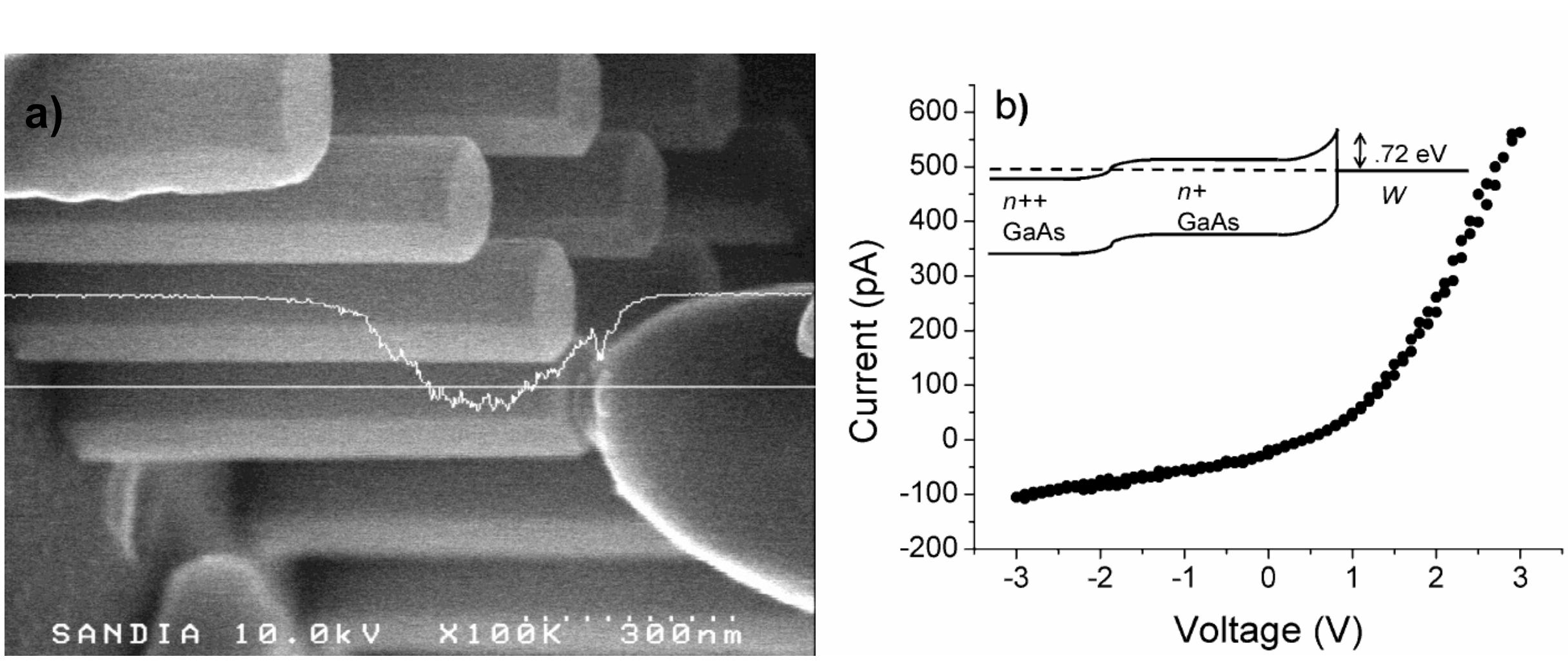

**Figure 3: a)** An EBIC profile (at zero applied bias) superposed on the secondary electron image. The straight line shows the trajectory of the e-beam scan along the length of the NW and the corresponding current profile indicates net current is collected only near the NW/tip interface. The acceleration energy and beam current were 10 keV and 10 pA respectively. **b)** A typical I-V curve obtained from an electroconductivity experiment. The electron beam was focused laterally in the center of the NW, 100 nm away from the W probe tip. The acceleration energy was 15 keV and the beam current, 10 pA. The creation of carriers reveals a low turn-on voltage and strong asymmetry. The expected band diagram at equilibrium is given in the inset.

I-V curves taken after electroconductivity experiments, with the e-beam off, reverse back to the trap-dominated transport behavior after several hours, which indicates the NWs were not damaged. While the electroconductivity diminished over time, it was still clearly identifiable even ~1 hour after e-beam exposure. This effect is not unique to e-beam excitation of carriers but is rather a consequence of the autolocalized centers we revealed above through analysis of the phonon-assisted tunneling transport. Persistent current has often been observed in the context of photoconductivity experiments and results from metastable states in which re-trapping is hindered by an energy barrier.[24]

In summary, we have shown that the semi-insulating electrical characteristic of catalyst-free MOCVD grown GaAs NWs is the result of trap-mediated carrier transport within the material. At intermediate electric fields Poole-Frenkel transport dominates crossing over to phonon-assisted tunneling at larger fields. By analyzing the phonon-assisted tunneling regime, we extracted the tunneling time, which enabled us to determine that the traps are autolocalized centers. Through electron-beam excitation techniques we confirmed the injection-limited nature of the system when traps are activated, thus revealing the nature of the contacts. The presence of traps in MOCVD-grown GaAs nanowires in conjunction with those previously observed in solution-grown material,[7] indicates that defects play a key role in this material system. Why would defects be more important in nanowires compared to thin films? One possibility is that surface states, which are known to be present at GaAs surfaces,[25] deplete the carriers in the nanowire thus enabling even modest trap densities to affect the transport.